\documentclass[11pt]{article}

\title{A Tool for Reconstructing Transit Vehicle Trajectories: A Case Study at IndyGo}
\author{%
  Ben O'Brien and Lewis J. Lehe\thanks{Corresponding author: \href{mailto:lehe@illinois.edu}{lehe@illinois.edu}.}\\
  Department of Civil and Environmental Engineering\\
  University of Illinois Urbana-Champaign
}
\date{\today}

\usepackage{enumitem}
\usepackage{ccaption}
\usepackage{newfloat}
\usepackage[fleqn]{amsmath}
\usepackage{graphicx}
\usepackage[margin=1in]{geometry}
\usepackage[numbers, square]{natbib}
\usepackage{xparse}
\usepackage{lastpage}
\usepackage{etoolbox} 
\usepackage[colorlinks=true,allcolors=blue,urlcolor=blue,citecolor=blue,linkcolor=blue]{hyperref}
\usepackage{xurl} 

\usepackage{soul}
\usepackage{tabularx}
\usepackage{algorithm}
\usepackage{algorithmic}
\usepackage{listings}
\usepackage{xcolor}
\usepackage{delimset}
\usepackage{booktabs}
\usepackage{stringstrings}
\usepackage{multirow}

\definecolor{codegreen}{rgb}{0,0.6,0}
\definecolor{codegray}{rgb}{0.5,0.5,0.5}
\definecolor{codepurple}{rgb}{0.58,0,0.82}
\definecolor{backcolour}{rgb}{0.95,0.95,0.92}
\lstdefinestyle{mystyle}{
    language=R,
    backgroundcolor=\color{backcolour},   
    commentstyle=\color{codegreen},
    keywordstyle=\color{magenta},
    deletekeywords={new,distance,deriv,sequence},
    numberstyle=\tiny\color{codegray},
    stringstyle=\color{codepurple},
    basicstyle=\ttfamily\footnotesize,
    breakatwhitespace=false,         
    breaklines=true,    captionpos=b,                    
    keepspaces=true,
    numbersep=5pt, showspaces=false,                
    showstringspaces=false, showtabs=false, tabsize=2}
\graphicspath{ {./images/} }
\DeclareFloatingEnvironment[
  fileext=los,
  listname={List of Videos},
  name=Video,
  placement=htbp
]{video}

\newcolumntype{Y}{>{\centering\arraybackslash}X}
\newcolumntype{W}{>{\centering\arraybackslash}p{1cm}}
\newcolumntype{Z}{>{\centering\arraybackslash}p{2cm}}

\begin{document}
\maketitle

\begin{abstract}
    Automatic vehicle location (AVL) data produced by transit vehicles is invaluable in performance studies, but turning raw AVL points into a detailed view of vehicle stop-and-gos is burdensome: the datasets are sparse, noisy, and prone to blunders. While recent research has explored methods of reconstructing trajectories describing the position of vehicles over time, the common techniques can be complex, and no open-source tools exist to help practitioners process the raw AVL. We fill this gap by proposing a thorough methodology for cleaning transit AVL data and providing an open-source R package, built on open data standards, to implement the workflow and reconstruct vehicle trajectories, allowing practitioners to easily formulate custom microscopic performance metrics. Using a large AVL dataset with over 3,000 trips from Indianapolis, Indiana, we demonstrate the workflow, evaluate the package's efficiency, and demonstrate the utility of the trajectories by estimating various traffic signal performance metrics. Finally, we use cross-validation to quantify the error in reconstructed trajectories at various polling frequencies. We find that the proposed data processing methodology and tool are efficient, requiring roughly 3 minutes of processing time on the large dataset, and error in position estimates is low (root mean square error under 10 meters at polling frequencies of 15 seconds). The estimated signal performance metrics can inform future evaluations of signal programming along the corridor. In sum, this paper serves as a practical guide and complementary toolbox for practitioners seeking to use transit AVL data to build a detailed view of vehicle stop-and-go cycles.
\end{abstract}


\newpage

\newpage
\section{Introduction}
\label{sec:intro}

As congestion is American cities worsens, public transit agencies and municipal streets departments across the country are increasingly investing in bus priority programs to improve the speed and reliability of transit service \citep{national_association_of_city_transportation_officials_move_2023}. Before an agency implements a delay mitigation treatment, however, they must diagnose causes of delay and unreliability in their system, and after a treatment is implemented, its impacts should be evaluated.

A common data source in route performance measurement is automatic vehicle location (AVL) data, repeated global positioning system (GPS) speed and location pings from transit vehicles \citep{coghlan_assigning_2019}. While these datasets can be rich, fully utilizing them is challenging: the datasets are large and noisy, and to our knowledge, no open-source tools exist to help practitioners analyze transit AVL data. Due to this, both commercial solutions and in-house public-sector tools \citep{california_department_of_transportation_california_2026} tend to provide practitioners with aggregated, macroscopic metrics, such as average stop-to-stop travel times. This can limit agencies' view of vehicles' individual stop-and-go cycles between stops and the delays these stop-and-gos cause. If a practitioner or researcher wanted to remedy this by applying one of many GPS analysis tools from other fields \citep{garcez_duarte_experimental_2025}, they would likely find the techniques ill-suited for transit AVL: transit vehicles have the advantage of progressing in one direction along a known route, eliminating the need for map-matching, but the limitation of low polling frequencies (often 15-30 seconds, versus 5 seconds or better in connected automobile research \citep{waddell_utilizing_2020}) relative to their stop-and-go cycles (both stop dwells and signal delays are often on the order of 15-30 seconds). As a result, common GPS smoothing and state estimation techniques can oversmooth real sources of delay in urban traffic \citep{robbennolt_comparative_2026,toledo_estimation_2007}.

\begin{figure}[ht]
    \centering
    \includegraphics[width=0.8\textwidth, keepaspectratio]{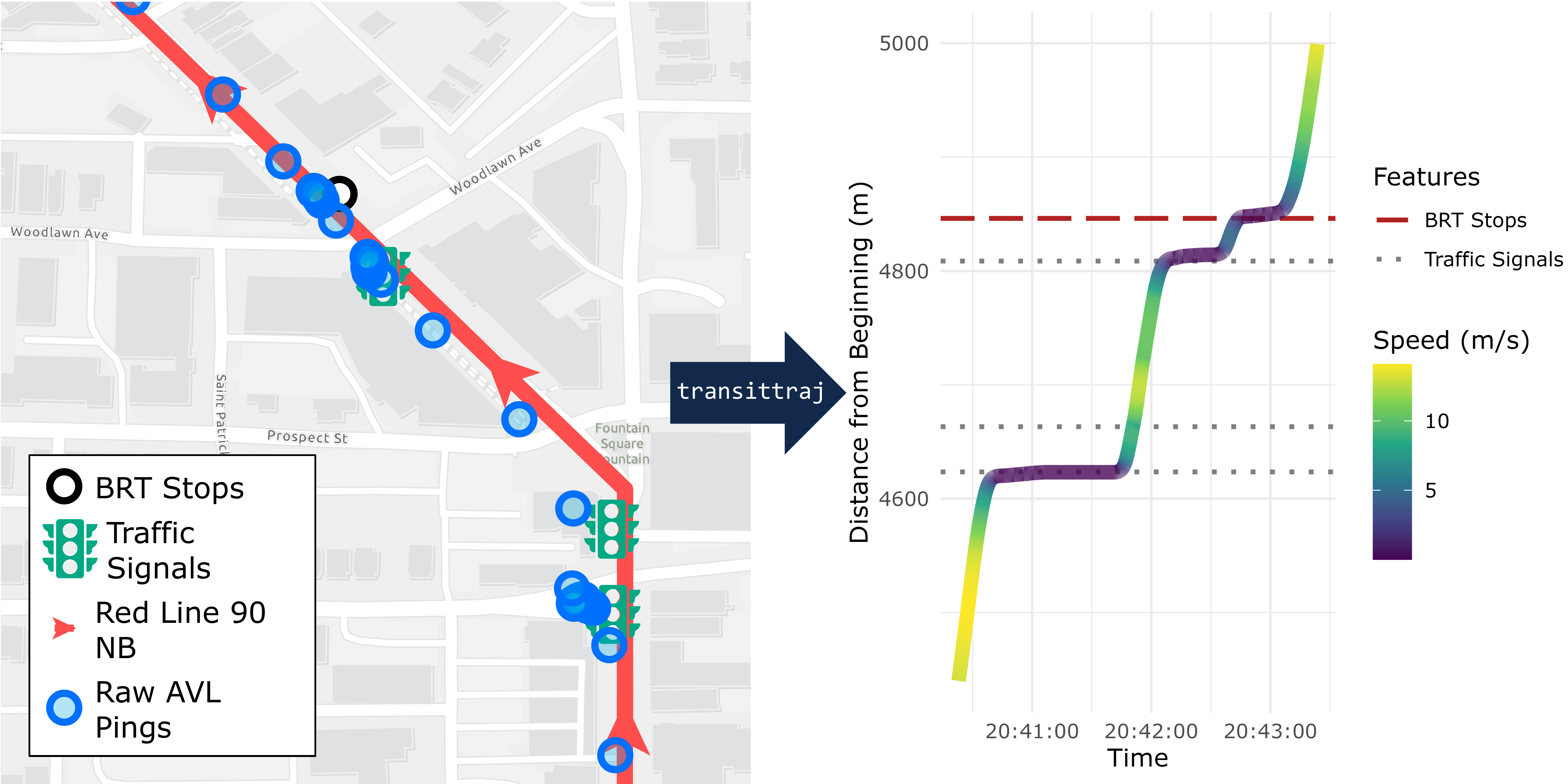}
    \caption{Transit vehicle trajectory (right) reconstructed from raw AVL (left).}
    \label{fig:traj}
\end{figure}

Recent research has has provided insight into techniques which can resolve the latter issue, primarily focusing on reconstructing vehicle \textit{trajectories} from AVL data \citep{huang_reconstructing_2023,robbennolt_comparative_2026}. A trajectory is a function describing the one-dimensional position (i.e., distance) of a transit vehicle on its route over time (Figure \ref{fig:traj}) and has four useful properties:

\begin{enumerate}
    \item The trajectory should be \textit{continuous}, not having any gaps;
    \item \textit{strictly monotonic}, with the vehicle's position always increasing with time;
    \item \textit{differentiable}, allowing the user to retrieve a vehicle's speed profile;
    \item and \textit{invertible}, allowing the user to retrieve the time a vehicle passed a specific point on its route.
\end{enumerate}
The first three attributes have been discussed by \citet{huang_reconstructing_2023}; we include the fourth, invertibility, to aid in the formulation of performance metrics.

Together, these four attributes make the trajectory function a versatile tool, giving practitioners a microscopic view of vehicles' speeds and stop-and-go cycles between scheduled stops. Despite the recent work surrounding transit trajectories, the reconstruction techniques proposed in the literature can be technically complex, and the data cleaning and wrangling required prior to reconstruction is under-discussed and burdensome. As a result, practical barriers remain for transit researchers and practitioners seeking to gain access to accurate and useful vehicle trajectory data.

To fill this gap, we present \texttt{transittraj}, an open-source \texttt{R} package for reconstructing transit vehicle trajectories from raw AVL data. \texttt{transittraj} relies on two open data standards: (1) the General Transit Feed Specification (GTFS), and (2) the Transit Integrated Data Exchange Specification (TIDES). While GTFS has become commonplace for storing static route and schedule information \citep{mobilitydata_general_2026}, TIDES is a relatively new standard for automated data archives, including AVL, and is finding adoption from agencies across the country \citep{mobilitydata_transit_2025}. \texttt{transittraj} is available through the Comprehensive \texttt{R} Archive Network (CRAN; \url{https://cran.r-project.org/web/packages/transittraj}), meeting the network's strong backend and user-facing standards. The package website (\url{https://utel-uiuc.github.io/transittraj/}) hosts extensive documentation, example code, and an example TIDES dataset. Together, these design choices make the tool accessible for practitioners and researchers, and for analysts using large language models (LLMs) to aid coding, \texttt{transittraj} represents a stable and robust human-developed foundation with ample example code for LLMs to learn from.


\texttt{transittraj} makes two novel contributions. First, we are the first to propose and describe, in detail, a thorough cleaning methodology for transit AVL data. We provide techniques to address both noise in urban GPS and the blunders commonly found in AVL datasets (such as deadheads logged to a revenue trip). Second, we are the first to present an accessible, open-source tool to implement these methodologies and common trajectory reconstruction techniques. To our knowledge, \texttt{transittraj} is one of the first packages designed to take advantage of the open TIDES standard. In sum, this paper serves as a practical guide, complemented by the \texttt{transittraj} toolbox, for practitioners and researchers seeking to use transit AVL data to understand vehicle stop-and-go cycles and formulate custom microscopic performance metrics.

This paper is structured as follows. In Section \ref{sec:workflow}, we present the package's architecture and discuss our proposed data cleaning and trajectory reconstruction methodology. Next, in Section \ref{sec:case_study}, we use \texttt{transittraj} on a large dataset from Indianapolis, Indiana, evaluating the package's efficiency and demonstrating its utility by formulating traffic signal performance metrics. In Section \ref{sec:robustness}, we quantify the error in the reconstructed Indianapolis trajectories using cross-validation. Finally, in Section \ref{sec:conclusion}, we summarize our findings and recommendations to practitioners and researchers.

\section{Package Architecture and Workflow}
\label{sec:workflow}

\begin{figure}[ht]
    \centering
    \includegraphics[width=\textwidth, keepaspectratio]{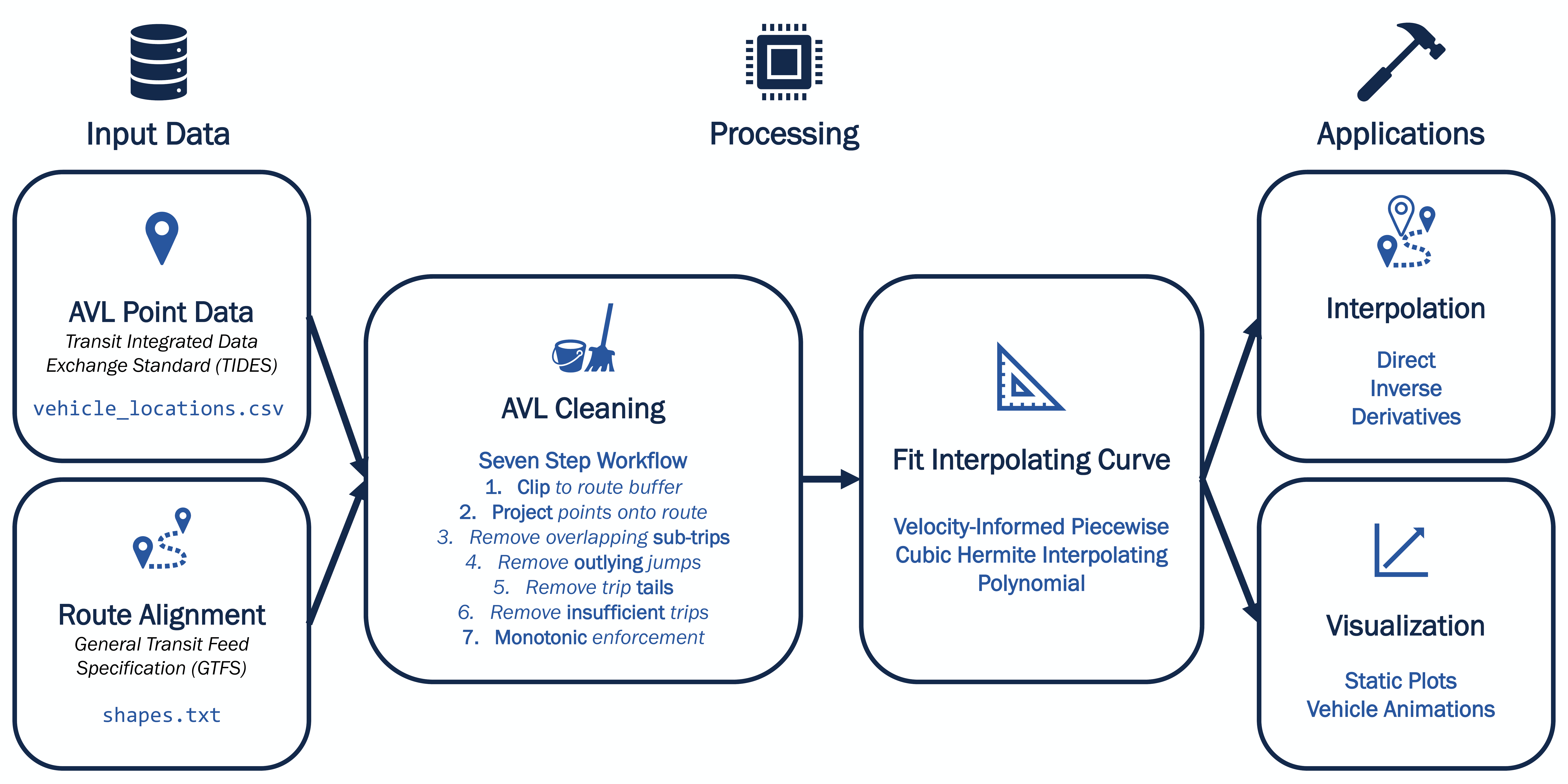}
    \caption{\texttt{transittraj} package architecture}
    \label{fig:arch}
\end{figure}

The architecture of \texttt{transittraj} is shown in Figure \ref{fig:arch}. The workflow begins with two types of input data, AVL points and the route alignment (Section \ref{sec:in_data}), followed by a seven-step cleaning workflow (Sections \ref{sec:spatial} through \ref{sec:mono}). Each cleaning step has a dedicated function with customizable tuning parameters. After cleaning, we fit a curve representing the trajectory of each trip (Section \ref{sec:fit}), which can be used for interpolation and visualization (Section \ref{sec:applications}). We implement the workflow in \texttt{R} because the language is highly accessible to both practitioners and researchers, and it allows us to take an efficient and readable ``tidy'' approach to data wrangling \citep{wickham_tidy_2014} through the \texttt{tidyverse} suite of packages \citep{wickham_welcome_2019}.

Many common trajectory reconstruction techniques risk oversmoothing the frequent and short stop-and-go cycles transit vehicles experience in urban traffic \citep{robbennolt_comparative_2026}. To maintain a clear view of these delays, our proposed workflow avoids smoothing techniques aimed at reducing low-level noise, and instead focuses on preserving the original location and speed data and interpolating between these points. Before interpolation can be performed, we correct and remove individual points suspected as errors or blunders, and remove trips with insufficient data. The \texttt{transittraj} workflow is intended to work with an arbitrary number of trips following the same shape (a single pattern of a single route traveling in a single direction). Each cleaning step is applied to each trip individually, and the final outcome of the workflow is a collection of functions, with each function describing the trajectory of one trip.

\begin{figure}[p]
    \centering
    \includegraphics[width=\textwidth, height=\textheight, keepaspectratio]{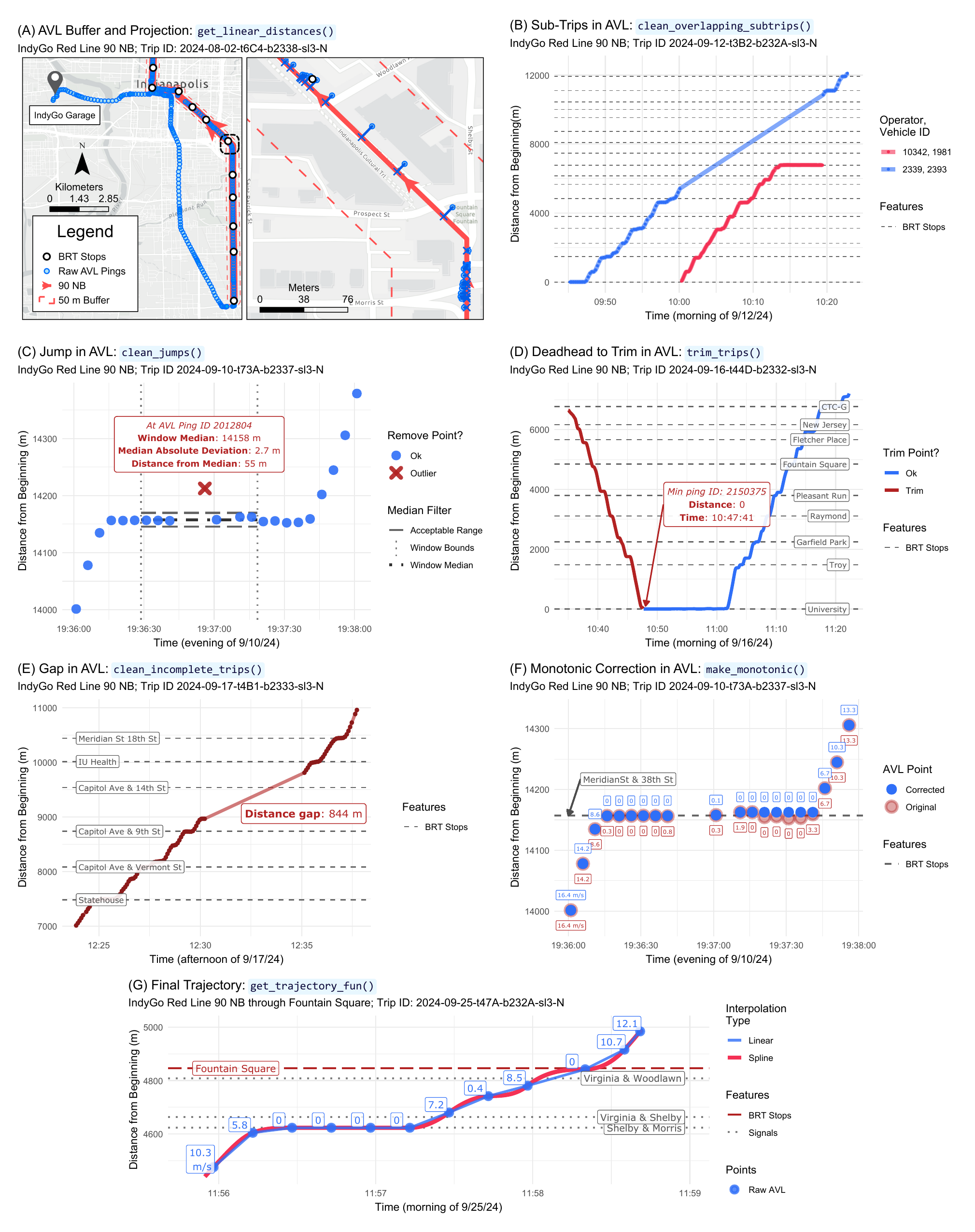}
    \caption{Examples of each step in the proposed AVL cleaning workflow.}
    \label{fig:cleaning}
\end{figure}

The following sub-sections describe each step of our proposed workflow and its algorithms. While we represent many of these algorithms using loops or element-wise operations for clarity, we implement them using vectorized calls to \texttt{tidyverse} functions wherever possible. Each step of the workflow is visualized in Figure \ref{fig:cleaning} using real AVL data from IndyGo (described in Section \ref{sec:case_study_data}). Finally, Section \ref{sec:applications} briefly describes \texttt{transittraj}'s tools for utilizing fit trajectory curves.

\subsection{Data Acquisition}
\label{sec:in_data}

\texttt{transittraj} requires two input data sources: (1) historic AVL data and (2) spatial route alignment data. First, AVL should take the form of a table matching the TIDES \texttt{vehicle\_locations.csv} schema \citep{mobilitydata_transit_2025}, and include longitude, latitude, timestamp, and trip ID information. Preferably, each point should also include a vehicle ID, operator ID, and vehicle speed. Second, route alignment data (the shape of the route) should come in the form of a GTFS \texttt{shapes.txt} file \citep{mobilitydata_general_2026}. \texttt{transittraj} includes validator functions to check that input data meets the required standards.

\subsection{Cleaning Steps 1 \& 2: Clip to Buffer \& Project onto Route}
\label{sec:spatial}

The \texttt{transittraj} workflow includes two geospatial steps, grouped into one function \texttt{get\_linear\_distances()}. The aim of these two steps is to turn two-dimensional latitude-longitude GPS points into one-dimensional distances along a trip's route. These steps are described in Algorithm \ref{alg:spatial} and implemented using the \texttt{R} package \texttt{geos}, a high-performance wrapper around the \texttt{C} library of the same name \citep{dunnington_geos_2026}.

\begin{algorithm}[ht]
    \centering
    \caption{Geospatial buffer and projection.}
    \label{alg:spatial}
    \begin{algorithmic}
        \REQUIRE $1 \times n$ vectors of GPS points, $\boldsymbol{P} = \left( \vec{\text{lon}}, \vec{\text{lat}} \right)$, on WGS84 ellipsoid
        \REQUIRE Linestring $\boldsymbol{R}$ representing the route alignment, in Euclidean coordinates
        \STATE \COMMENT{\textit{Step 1: Clip points to route buffer}}
        \STATE Project GPS to Euclidean coordinates: $\boldsymbol{P} \left( \vec{\text{lon}}, \vec{\text{lat}}  \right) \rightarrow \boldsymbol{P} \left( \vec{E}, \vec{N} \right)$
        \STATE Construct polygon $\boldsymbol{B}$ with buffer of radius $b$ around alignment $\boldsymbol{R}$
        \STATE Clip points to buffer: $\boldsymbol{P}' \leftarrow \boldsymbol{P} \cap \boldsymbol{B}$
        \STATE \COMMENT{\textit{Step 2: Project points onto route}}
        \STATE Initialize vector of one-dimensional distances $\vec{x} = \left( \right)$
        \FOR{$\vec{p} \in \boldsymbol{P}'$}
            \STATE $x_i \leftarrow \text{proj} \left( p \text{ onto } \boldsymbol{R} \right)$
        \ENDFOR
        \ENSURE Vectors of one-dimensional distances along the alignment route $\vec{x}$
    \end{algorithmic}
\end{algorithm}

In step 1, a buffer of distance $b$ is created around the route alignment, and only points within this buffer are kept. This removes both outliers lateral to the route and off-route deadheading. For example, Figure \ref{fig:cleaning}(A) shows a trip deadheading from the bus garage before reaching its route. Next, in step 2, each point is projected onto the route alignment (``snapped'' to the nearest point on the route shape), returning a distance from the alignment's beginning. \citet{punzo_assessment_2011} show that, when the trajectory of interest is only the longitudinal position along a known path, projection eliminates biases in distance-traveled estimates due to random GPS noise. This is an approach similar to \citet{huang_reconstructing_2023}, who use OpenStreetMap features rather than GTFS shapes. Because GTFS shapes are directional lines, with the shape's beginning at the trip beginnings, the distance values estimated by \texttt{transittraj} start near zero and increase as the trip progresses.

For the practitioner, there is one decision variable: the distance buffer $b$. The user should also choose an appropriate coordinate system. The output distance vector $\vec{x}$ will be in units of the coordinate system used, so a Euclidean projection will aid in interpretability.

\subsection{Cleaning Step 3: Remove Overlapping Sub-trips}

In many AVL systems, it may be possible for multiple vehicles or operators to register pings under the same trip ID. In some cases -- such as a mid-route operator change -- this is expected and acceptable. Other times, however, these unique trip-vehicle-operator combinations (``sub-trips'') overlap, making it difficult to know which sub-trip actually represents revenue service. Step 3, implemented via \texttt{clean\_overlapping\_subtrips()}, identifies these scenarios and removes these trips. Step 3 is described in Algorithm \ref{alg:subtrips}.

\begin{algorithm}[ht]
    \centering
    \caption{Sub-trip identification and removal.}
    \label{alg:subtrips}
    \begin{algorithmic}
        \REQUIRE $1 \times n$ vectors of points for one trip ID, ordered by time, $\boldsymbol{P} = \left( \vec{x}, \vec{t}, \vec{\text{veh}}, \vec{\text{op}} \right)$, where $x$ is the point's one-dimensional distance, $t$ is the point's timestamp, $\text{veh}$ is the point's vehicle ID, and $\text{op}$ is the point's operator ID
        \STATE Define sub-trip IDs $\vec{s} = \text{concat} \left( \vec{\text{veh}}, \vec{\text{op}} \right)$
        \STATE Initialize $\boldsymbol{T} \leftarrow \left( \right)$ to store sub-trip time ranges, each a two-element vector
        \STATE \COMMENT{\textit{For each sub-trip ID, retrive its time range}}
        \FOR{$s \in \text{unique} \left( \vec{s} \right)$}
            \STATE $\vec{t}_s = t$ where $\vec{s} = s$
            \STATE $\vec{T}_s = \left( \text{min}\left(\vec{t}_s \right), \text{max}\left(\vec{t}_s \right) \right)$
        \ENDFOR
        \STATE \COMMENT{\textit{For each sub-trip, check if its time range overlaps with any other subtrip. Accomplished efficiently with \texttt{R} package \texttt{ivs}, but represented here as nested for-loops.}}
        \STATE $\boldsymbol{T} \leftarrow \text{arrange} \left( \boldsymbol{T} \text{ by increasing } \min{\vec{T}} \right)$
        \FOR{$\vec{T}_i \in \boldsymbol{T}$}
            \FOR{$\vec{T}_j \in \boldsymbol{T}$ where $j > i$}
                \IF{$\max{\vec{T}_i} > \min{\vec{T}_j}$}
                    \STATE Remove trip
                \ENDIF
            \ENDFOR
        \ENDFOR
        \ENSURE $\boldsymbol{P}$ if trip is acceptable
    \end{algorithmic}
\end{algorithm}

An example is shown in Figure \ref{fig:cleaning}(B); here, operator 10324, driving vehicle 1981, logs into the trip ID at the same time as operator 2339, driving vehicle 2393. For the practitioner, there is one main decision variable: should trips with non-overlapping sub-trips be removed, or only those with overlap?

\subsection{Cleaning Step 4: Remove Outlying Jumps}

In step 4, implemented via \texttt{clean\_jumps()}, \texttt{transittraj} attempts to identify and remove outlying observations in the dataset. This is accomplished via a type of sliding-window median filter known as a \textit{Hampel filter}. These filters construct a window of width $k$ around each point and remove the point if it lies beyond some acceptable distance, in median absolute deviations (MADs), from the window median.

\begin{algorithm}[ht]
    \centering
    \caption{Standard Hampel filter.}
    \label{alg:HF}
    \begin{algorithmic}
        \REQUIRE $1 \times n$ vectors of points for one trip ID, ordered by time, $\boldsymbol{P} = \left( \vec{x}, \vec{t} \right)$, where $x$ is the point's one-dimensional distance, and $t$ is the point's timestamp
        \REQUIRE Hampel's $t$ outlier cutoff, and two-sided neighborhood width $k$
        \STATE Initialize $\boldsymbol{P}' \leftarrow \left( \right)$, cleaned vector of position and time points
        \FOR{$i = k/2 + 1, k/2 + 2, ..., n - k/2$}
            \STATE $\vec{W}_i = \left( x_{i - k/2}, ..., x_i, ..., x_{i + k/2} \right), \quad m_i = \text{median} \left( \vec{W}_i \right), \quad \text{MAD}_i = \text{median} \left( \vec{W}_i - m_i \right)$
            \IF{$\abs{x_i - m_i} > t s \text{MAD}_i \quad \text{and} \quad \text{MAD}_i \ne 0$}
                \STATE Remove observation
            \ELSE
                \STATE $\vec{p}'_i \leftarrow \left( x_i, t_i \right)$
            \ENDIF
        \ENDFOR
        \ENSURE $\boldsymbol{P}'$ of length $\le n$ of cleaned position and time points
    \end{algorithmic}
\end{algorithm}

The Hampel filter is described in Algorithm \ref{alg:HF}. Here, $s$ is a conversion factor making the MAD an unbiased estimate of standard deviation, with $s = 1.483$ for normally distributed data, and $t$ is a tuning parameter, describing the allowable number of standard deviations a point may be from its window's median. A cutoff of $t = 3$ is a common rule of thumb \citep{pearson_generalized_2016,garcez_duarte_experimental_2025}, and is \texttt{transittraj}'s default. An example window at one point is shown in Figure \ref{fig:cleaning}(C). Here, the point's window has an MAD of 2.7 meters and a maximum allowable distance of $t s \text{MAD} = 3 \times 1.483 \times 2.7 = 12.0$ meters; because the point is 55 meters from the window median, it was identified as an outlier and removed.

While Hampel filters are used by open-source trajectory reconstruction tools for other fields \citep{garcez_duarte_experimental_2025}, to our knowledge they have not been applied to transit AVL data. Hampel and median filters are well-suited to these applications, though, for two reasons. First, they require minimal tuning and are easy to understand, making \texttt{transittraj} more accessible to practitioners. Second, median filters focus on removing instantaneous jumps, rather than smoothing low-level noise \citep{pearson_generalized_2016}, unlike many other common outlier replacement strategies \citep{garcez_duarte_experimental_2025}, reducing the risk of oversmoothing.

Despite their advantages, Hampel filters have two limitations. The first limitation is that a complete window cannot be formed at the beginning and end of a vector of observations, hurting the filter's performance \citep{pearson_generalized_2016}. This is generally acceptable for transit trajectories, as our goal is not to develop a detailed view of terminal operations. By default, \texttt{clean\_jumps()} will skip trip tails before a window can be formed. The second limitation is implosion sequences, which occur when more than half of a window's points have identical values, resulting in $\text{MAD} = 0$ and a guarantee that a point will be removed \citep{pearson_generalized_2016}. Due to random noise in GPS readings, these are uncommon in transit vehicle trajectories. By default, \texttt{clean\_jumps()} will identify and skip points in an implosion sequence.

For the practitioner, there are two primary decision variables when using \texttt{clean\_jumps()}: the neighborhood width $k$ (default $k = 7$ observations), and the Hampel cutoff $t$ (default $t = 3$ standard deviations). If desired, the practitioner can instead use raw distance as the median filter cutoff (for example, remove points more than 50 meters from their window median).

\subsection{Cleaning Step 5: Remove Trip Tails}

In step 1, some deadhead points were removed by keeping only AVL pings which were near the route. However, pings received when a vehicle was deadheading along its route may still be in the dataset. Step 5 attempts to identify and remove these pings. To accomplish this, the function \texttt{trim\_tails()} identifies the points with the minimum and maximum distance; points before the minimum or after the maximum are removed (Algorithm \ref{alg:trim}). For example, in Figure \ref{fig:cleaning}(D), the vehicle travels backwards along the route before holding at the beginning terminal, then starting the trip in the correct direction of travel. For the practitioner, the only decision variable is whether to trim each trip's beginning, ending, or both (the default).

\begin{algorithm}[ht]
    \centering
    \caption{Trimming trip tails.}
    \label{alg:trim}
    \begin{algorithmic}
        \REQUIRE $1 \times n$ vectors of points for one trip ID, ordered by time, $\boldsymbol{P} = \left( \vec{x}, \vec{t} \right)$, where $x$ is the point's one-dimensional distance, and $t$ is the point's timestamp
        \STATE $j = \underset{i \in \left[1, n\right]}{\text{argmin }} \vec{x}, \quad k = \underset{i \in \left[1, n\right]}{\text{argmax }} \vec{x}$
        \IF{$k \le j$}
            \STATE Remove trip
        \ENDIF
        \STATE $\boldsymbol{P}' = \leftarrow \left( \left( x_j, x_{j + 1}, ..., x_{k - 1}, x_k \right), \left( t_j, t_{j + 1}, ..., t_{k - 1}, t_k \right)\right) $
        \ENSURE $\boldsymbol{P}'$ of length $k - j + 1 \le n$ of trip points without tails, if trip is acceptable
    \end{algorithmic}
\end{algorithm}

\subsection{Cleaning Step 6: Remove Insufficient Trips}

AVL feeds may, at times, miss a period of pings, and some trips may only contain a few recorded points. Additionally, previous outlier removal (step 4) and tail trimming (step 5) may result in gaps in trajectories or unusually short trips. As a result, not all trips in the dataset will have the amount of data necessary to formulate accurate performance metrics. Step 6, implemented via \texttt{clean\_incomplete\_trips()} and described in Algorithm \ref{alg:insuf}, removes trips not meeting user-defined distance and time standards. For example, the trip in Figure \ref{fig:cleaning}(E) has a large, unexplained gap of roughly 840 meters over 5 minutes; depending on how the practitioner is using the final trajectories, it may be unreasonable to interpolate over this gap.

The practitioner has two types of decision variables: minimum trip durations (in terms of distance $D_\text{min}$ or time $T_\text{min}$), and maximum allowable gaps (in terms of distance $\Delta^x_\text{max}$ or time $\Delta^t_\text{max}$). Trips not meeting these requirements will be removed, in entirety, from the dataset.

\begin{algorithm}[ht]
    \centering
    \caption{Removing insufficient trips.}
    \label{alg:insuf}
    \begin{algorithmic}
        \REQUIRE $1 \times n$ vectors of points for one trip ID, ordered by time, $\boldsymbol{P} = \left( \vec{x}, \vec{t} \right)$, where $x$ is the point's one-dimensional distance, and $t$ is the point's timestamp
        \REQUIRE Minimum allowable trip distance and duration ($D_\text{min}$ and $T_\text{min}$), and maximum allowable distance and time gaps ($\Delta^x_\text{max}$ and $\Delta^t_\text{max}$)
        \STATE $D = \max{\vec{x}} - \min{\vec{x}}, \quad T = \max{\vec{t}} - \min{\vec{t}}$
        \STATE $\Delta^x = \max{ \left( x_2 - x_1, x_3 - x_2, ..., x_{n} - x_{n - 1} \right) }, \quad \Delta^t = \max{ \left( t_2 - t_1, t_3 - t_2, ..., t_{n} - t_{n - 1} \right) }$
        \IF{$D < D_\text{min} \quad \text{or} \quad T < T_\text{min} \quad \text{or} \quad \Delta^x > \Delta^x_\text{max} \quad \text{or} \quad \Delta^t > \Delta^t_\text{max} $}
            \STATE Remove trip
        \ENDIF
        \ENSURE $\boldsymbol{P}$ if trip is acceptable
    \end{algorithmic}
\end{algorithm}

\subsection{Cleaning Step 7: Monotonic Enforcement}
\label{sec:mono}

The seventh and final data cleaning step, implemented via \texttt{make\_monotonic()}, enforces strict monotonicity onto a vector of times, positions, and speeds. The monotonic enforcement algorithm, described in Algorithm \ref{alg:ME}, is similar to what is proposed by \citet{robbennolt_comparative_2026}. The recursive algorithm returns a set of position, time, and speed points in which the distance of every point is greater than the point before it, and the speed of every point is greater than zero and satisfies the Fritsch-Carlson constraints \citep{fritsch_monotone_1980} for a monotonic cubic spline.

\begin{algorithm}[ht]
    \centering
    \caption{Strict monotonic enforcement of position and speed.}
    \label{alg:ME}
    \begin{algorithmic}
        \REQUIRE $1 \times n$ vectors of points for one trip ID, ordered by time, $\boldsymbol{P} = \left( \vec{x}, \vec{t}, \vec{v} \right)$, where $x$ is the point's one-dimensional distance, $t$ is the point's timestamp, and $v$ is the point's speed
        \REQUIRE Distance perturbation to add $\varepsilon$
        \STATE Initialize $\vec{s} \leftarrow \left(   \right)$ to track whether each point's distance was adjusted
        \STATE \COMMENT{\textit{Apply initial monotonic correction to positions \& zero-speeds}}
        \STATE Initialize $\boldsymbol{P}' \leftarrow \boldsymbol{P}$, vectors of to-be-corrected points
        \FOR{$i = 2, 3, ..., n$}
            \IF{$v_i = 0$}
                \STATE $v_i' \leftarrow \varepsilon / \left( t_i - t_{i - 1} \right) $
            \ENDIF
            \IF{$x_i \le x_{i - 1}$}
                \STATE $x_i' \leftarrow x_{i - 1} + \varepsilon, \quad s_i \leftarrow 1$
            \ELSE
                \STATE $s_i \leftarrow 0$
            \ENDIF
        \ENDFOR
        \STATE \COMMENT{\textit{Adjust speeds according to new positions}}
        \FOR{$i = 2, 3, ..., n - 1$}
            \IF{$s_i = 1$}
                \STATE $v_i' \leftarrow \max{ \left[ v_i', \quad \frac{x_{i + 1}' - x_{i - 1}'}{t_{i + 1}' - t_{i - 1}'} \right] } $
            \ENDIF
        \ENDFOR
        \STATE \COMMENT{\textit{Apply Fritsch-Carlson monotonic speed conditions}}
        \FOR{$i = 1, 2, ..., n - 1$}
            \STATE $\delta_i = \frac{x_{i + 1}' - x_i'}{t_{i + 1}' - t_i'}, \quad \alpha_i = v_i' / \delta_i, \quad \beta_i = v_{i + 1}' / \delta_i, \quad \tau_i = 3 / \sqrt{\alpha_i^2 + \beta_i^2}$
            \IF{$\alpha_i^2 + \beta_i^2 > 9$}
                \STATE $v_i' \leftarrow \tau_i \alpha_i \delta_i, \quad v_{i + 1}' \leftarrow \tau_i \beta_i \delta_i$
            \ENDIF
        \ENDFOR
        \ENSURE $\boldsymbol{P}'$ of length $n$ of monotonic distance, time, and speed points
    \end{algorithmic}
\end{algorithm}

Algorithm \ref{alg:ME} first corrects speed values observed to never be exactly zero, increasing them relative to a user-defined distance perturbation $\varepsilon$. Then, if any point has a distance less than the point behind it, it is pulled up to that point plus $\varepsilon$. The speed at this point is then reset to the maximum of observed speed and the speed implied by the corrected distance and time observations on either side. Finally, the Fritsch-Carlson slope conditions, which guarantee a fit cubic spline will be monotonic \citep{fritsch_monotone_1980}, are used to correct the speeds. The Fritsch-Carlson conditions are discussed more thoroughly in Section \ref{sec:fit}. An example is shown in Figure \ref{fig:cleaning}(F), where the AVL pings drift backwards slightly while the vehicle holds at a BRT stop. The function pulls these points up and makes small corrections to the observed speeds during this stop.

For the practitioner, there is one primary decision variable when using \texttt{make\_monotonic()}: the added distance error $\varepsilon$. To achieve strict monotonicity (and thus invertibility) in the final trajectory, we recommend the practitioner choose some small value in the units of their distance vector (e.g., $\varepsilon = 0.001 \text{ m} = 1 \text{ mm}$).

\subsection{Fit Interpolating Curve}
\label{sec:fit}

Once cleaning is complete, an interpolating curve can be fit to the strictly monotonic data. \citet{huang_reconstructing_2023} and \citet{robbennolt_comparative_2026} provide thorough evaluations and discussions of interpolation techniques appropriate for transit vehicles. \citet{huang_reconstructing_2023} recommends, in the absence of observed speeds, using a combination of local regression smoothing and cubic spline interpolation to produce realistic trajectories. \citet{robbennolt_comparative_2026} find that when observed speeds are available, however, smoothing techniques (including local regression) can harm the accuracy of the fit trajectory. Instead, using observed speeds to fit an unsmoothed cubic spline results in less prediction error, more realistic acceleration curves, and better alignment with known dwell times.

Following \citet{robbennolt_comparative_2026}, \texttt{transittraj} fits a \textit{velocity-informed piecewise cubic interpolating polynomial with monotonic enforcement} (VCHIP-ME) to each trip's trajectory. VCHIP-ME is an adaptation of \citet{fritsch_monotone_1980}'s monotonic piecewise cubic interpolating polynomial (PCHIP) technique. Under PCHIP, for some new time $t$ lying between observed points $\vec{p}_i = \left(x_i, t_i, v_i \right)$ and $\vec{p}_{i + 1} = \left(x_{i + 1}, t_{i + 1}, v_{i + 1} \right)$, the cubic interpolating function will be:

\begin{equation}
    \centering
    \begin{aligned}
        x(t) &= x_i h_{00}(l) + \Delta_i v_i h_{10}(l) + x_{i + 1} h_{01}(l) + \Delta_i v_{i + 1} h_{11}(l) \\
        \text{Where: } & \\
        \Delta_i &= t_{i + 1} - t_i, \quad l = \left(t - t_i \right) / \Delta_i \\
        h_{00}(l) &= 2 l^3 - 3 l^2 + 1, \quad h_{10}(l) = l^3 - 2 l^2 + l, \quad h_{01}(l) = -2 l^3 + 3 l^2, \quad h_{11}(l) = l^3 - l^2
    \end{aligned}
\end{equation}

\citet{fritsch_monotone_1980} show that, if the original positions $\vec{x}$ are strictly increasing and the slopes $\vec{v}$ satisfy the conditions applied in third part of Algorithm \ref{alg:ME}, the PCHIP function is guaranteed to be strictly monotonic. The application of PCHIP to observed speeds, corrected to meet the Fritsch-Carlson constraints, is known as \textit{velocity-informed PCHIP} (VCHIP), and the correction of ``backtracking'' position values is known as \textit{monotonic enforcement} (ME). Together, cleaning step 7 (via Algorithm \ref{alg:ME}) and the fitting of a PCHIP function represent velocity-informed PCHIP with monotonic enforcement (VCHIP-ME). The VCHIP-ME technique yields a trajectory meeting the four requirements described in Section \ref{sec:intro}: the function is guaranteed to be continuous and monotonic, it is a polynomial and thus differentiable, and knowing that the first two requirements are satisfied, its inverse can easily be found numerically.

An example VCHIP-ME spline is compared to linear interpolation in Figure \ref{fig:cleaning}(G). This vehicle makes at least three stop-and-go cycles while traversing this corridor of dense signals and stations; the second two are short, lasting roughly the duration of the vehicle's polling frequency. Because of this, a simple, speed-ignorant linear interpolation misses the latter two stops and shows the vehicle traversing the corridor at a constant speed. The VCHIP-ME spline, however, uses the recorded near-zero speeds during these stops to draw all three stop-and-go cycles.

With cleaned AVL point data, users can easily fit interpolating curves using the function \texttt{get\_trajectory\_fun()}. For the practitioner, there are two main decision variables. First is the method of interpolation. While VCHIP-ME is recommended and is the default, \texttt{transittraj} also supports simple linear interpolation and speed-ignorant monotonic PCHIP if observed speeds are not available. Second, the user must select a tolerance, in distance units, for the numeric inverse function.

\subsection{Using Trajectories}
\label{sec:applications}

When the user applies \texttt{get\_trajectory\_fun()} to fit interpolating curves to each trip (Section \ref{sec:fit}), \texttt{transittraj} stores and returns the fit curves within a special object class. The object also contains information about each trip's time and distance range, as well as the settings used to fit the curves. \texttt{transittraj} supports two primary uses of the trajectory object: prediction and visualization. Both utilize class-specific methods for common generic functions to make these applications as accessible as possible to experienced and novice \texttt{R} users alike. Together, these applications provide practitioners and researchers with a versatile yet simple interface for using fit trajectory splines.

First, prediction allows the practitioner to interpolate new time, distance, and speed points along the trajectory without interacting directly with the cubic interpolating polynomial. The proposed trajectory object class has a \texttt{predict()} method which can use the interpolating polynomials in three main ways: first, \textit{direct interpolation} takes in new times and retrieves the vehicles' distances or speeds at these times; second, \textit{inverse interpolation} takes in new distances and retrieves the time at which each vehicle crossed these distance; finally, \textit{bounded interpolation} takes in distance bounds, retrieving time-distance-speed points at a desired temporal resolution between these bounds. The \texttt{predict()} method will return the interpolated values for all trips whose observed range crosses the desired distance or time range. Together, these three prediction options allow the practitioner to formulate and calculate a wide array of performance metrics with a single function call. Some example applications are explored in Section \ref{sec:case_study}.

The second application of trajectories is visualization, a powerful tool for both external stakeholder communication and internal diagnostics of vehicle performance concerns. \texttt{transittraj} supports two main types of visualizations: static trajectory plots, similar to Figure \ref{fig:traj}, through a dedicated \texttt{plot()} method; and vehicle animations, showing trips moving through space or a straight-line route, through \texttt{plot\_animated\_map()} and \texttt{plot\_animated\_line()}, respectively.

\section{Case Study: Traffic Signal Performance Measures at IndyGo}
\label{sec:case_study}

Indianapolis, Indiana is a large Midwestern city served by the Indianapolis Public Transportation Corporation (IndyGo), who operates two bus rapid transit (BRT) lines through the city. As with many BRT operators, IndyGo is concerned with measuring and mitigating delays due to traffic signals. In this section, we use \texttt{transittraj} to estimate various signal performance metrics on one of IndyGo's BRT routes, the Red Line (Route 90). Our goals here are twofold: first, to demonstrate the package's efficiency on a large, real-world dataset; and second, to demonstrate how \texttt{transittraj} can use trajectories to formulate useful microscopic performance metrics. The following sub-sections describe our data, methods, and results. While we do not have permission to share the raw data, we have made available all code used to process the AVL data and calculate the performance metrics through a supplementary GitHub repository (\url{https://github.com/UTEL-UIUC/transittraj_paper-supp}).

\subsection{Background}

\begin{figure}[ht]
    \centering
    \includegraphics[width=0.5\textwidth, keepaspectratio]{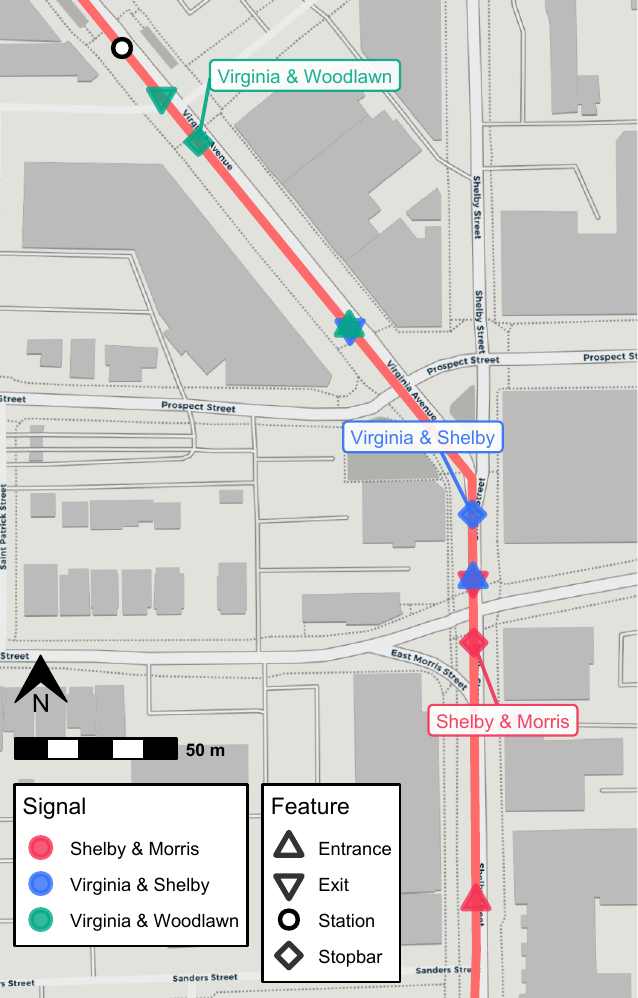}
    \caption{Map of northbound Fountain Square station, signal stopbars, and signal entrances and exits.}
    \label{fig:fnt_sqr}
\end{figure}

The Red Line is a 28-station, 21-kilometer route running north-south through both mixed traffic and dedicated center- and side-running bus lanes \citep{indygo_indygo_2021}. All traffic signals along the route are equipped with transit signal priority (TSP). We analyze signal performance in the northbound direction through a corridor in south Indianapolis named Fountain Square, a set of three signals within the span of 180 meters immediately followed by one station (Figure \ref{fig:fnt_sqr}).

We focus on three signal performance measures intended to represent both travel time and corridor progression. First, we estimate \textit{control delay} as the excess travel time each trip experiences beyond an intersection's free-flow time. Second, we estimate \textit{arrival-on-green} (AoG), the portion of trips which reach the signal while it is green in the direction of travel. Third, we estimate \textit{split failures} (SF), the portion of trips which do not clear the signal in one complete cycle. These three metrics are becoming a focus for practitioners seeking to understand the performance of TSP systems \citep{jackson_use_2024,coghlan_assigning_2019}. While these metrics are most often calculated using signal controller logs \citep{jackson_use_2024,nevers_nchrp_2020}, we here estimate them using reconstructed trajectories by identifying stop-and-go cycles at each signal. Following \citet{jackson_use_2024}, our focus is on signal performance from the perspective of transit vehicles, and we do not attempt to make statements about signal performance for the general traffic flow.

\subsection{Data}
\label{sec:case_study_data}

\begin{figure}[ht]
    \centering
    \includegraphics[width=\textwidth, keepaspectratio]{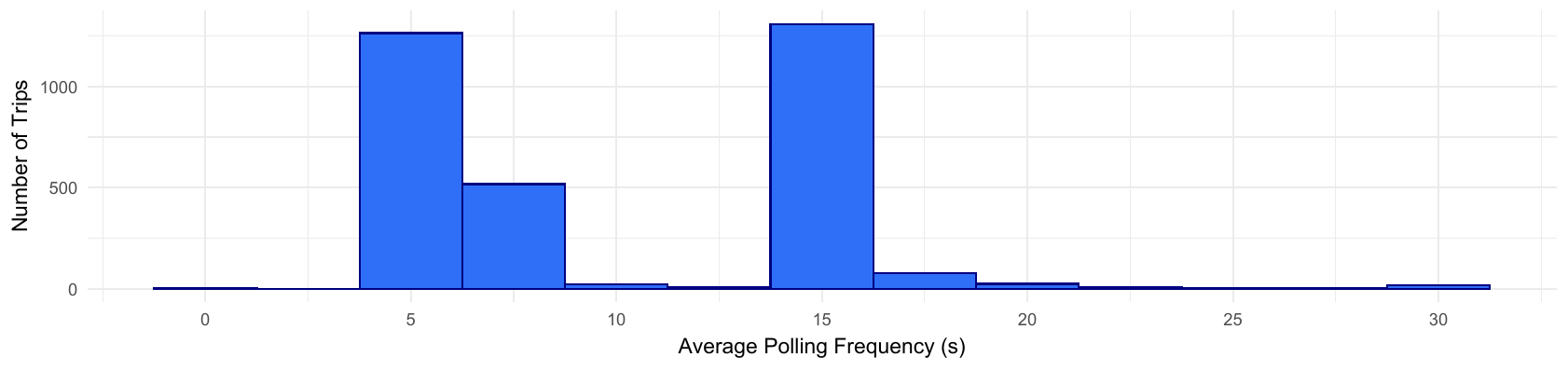}
    \caption{Histogram of average IndyGo AVL polling frequencies by trip.}
    \label{fig:polling_hist}
\end{figure}

Two types of data are required for this case study: (1) AVL pings and (2) geospatial feature data. First, AVL points are polled from IndyGo's BRT vehicles every 5 or 15 seconds, depending on the hardware equipped on each vehicle (Figure \ref{fig:polling_hist}). These pings include trip, vehicle, and operator IDs, in addition to latitude, longitude, speed, and timestamp data. IndyGo provided us with all northbound Red Line AVL pings from weekdays between November 1 and December 31, 2024. In total, the original dataset included 1,707,095 points across 3,256 unique trips with an average polling frequency of 7.99 seconds.

Second, spatial feature data included the northbound route alignment, stop locations, and signal stopbar locations. We retrieved alignment and stop location data from a historic version of IndyGo's GTFS accessed via Transitland \citep{transitland_transitland_2024}. We used the northbound Red Line route alignment stored in the feed's \texttt{shapes.txt} file, as described in Section \ref{sec:in_data}, and stop locations from the \texttt{stops.txt} file. We then identified signal stopbar locations (the thick, white pavement marking indicating where drivers should stop ahead of an intersection) for each direction of travel through Fountain Square using publicly-available satellite imagery. To understand how vehicle movements are influenced by the signals, we defined an ``entrance'' and ``exit'' point for each signal in the northbound travel direction (Figure \ref{fig:fnt_sqr}). The exit of each signal was taken as the southbound stopbar's location. For Virginia \& Woodlawn and Virginia \& Shelby, the entrance was taken as the upstream signal's exit, and for Shelby \& Morris, the entrance was taken as the point 80 meters upstream of the stopbar. The goal was the capture the entire deceleration and acceleration curves caused by each signal, but not the influence other signals, driveways, or the Fountain Square station.

We transformed all spatial feature data and AVL pings from their original WGS84 ellipsoid to the WGS84 Universal Transverse Mercator Zone 16N Euclidean plane. Finally, all stop and stopbar locations were snapped to the route alignment and converted to one-dimensional distances from the start of the route.

\subsection{Methods}

\subsubsection{AVL Cleaning}
\label{sec:case_study_methods}

We cleaned the raw AVL data using the complete \texttt{transittraj} data cleaning methodology using all default and recommended parameters described in Section \ref{sec:workflow}. In step 1, a buffer radius of 50 meters is used to remove points lying far from the route alignment. Following step 2 (spatial projection), we exclude all points in the northernmost 500 meters of the route. This is because, at the northern terminal, the route shape loops over itself and vehicles often layover on side streets or parking lots adjacent to the route, making this data unusually messy and difficult to parse. Because our goal, and the goal of \texttt{transittraj} as a whole, is not to understand terminal operations, we remove the 70,250 AVL pings at this terminal. In step 6 (removing insufficient trips), only trips with a minimum distance of 500 meters and duration of 90 seconds are kept, and trips with a gap larger than 800 meters are removed. Finally, a VCHIP-ME spline is fit for each trip, with a numeric inverse tolerance of 0.01 meters.

To understand the efficiency of \texttt{transittraj}, we use \texttt{R}'s \texttt{bench} package \citep{hester_bench_2025} to repeat each cleaning function call 20 times and measure the processing time required by each iteration. We here report the median time required for each function, as benchmarking results are often right-skewed \citep[Chapter 23]{wickham_advanced_2019}. While users' experiences will depend on the hardware they are using, this gives them a rough idea of the performance to expect and the relative time required for each step. We ran this workflow on a standard consumer laptop with a 14-core processor and 16 gigabytes of RAM.

\subsubsection{Signal Performance Measures}

With the reconstructed trajectories and intersection geometries, the \texttt{transittraj} \texttt{predict()} method can be used to find the times each trip enters and exits each signal as shown in Code Block \ref{lst:tt}. With entrance $t_i^{s,\text{in}}$ and exit $t_i^{s,\text{out}}$ times for each trip $i$ at each signal $s$, we calculate the travel times $TT^s_i$ as shown in Equation \ref{eq:tt}. We approximate the free-flow travel time as the 5th percentile of all observed travel times at each signal, then subtract this from the observed travel time to estimate signal control delay.

\begin{equation}
    \label{eq:tt}
    \begin{aligned}
        TT^s_i &= t^{s, \text{out}}_i - t^{s, \text{in}}_i \\
        \text{Delay}^s_i &= \max{\left[ 0, \quad TT^s_i - \text{quantile} \left( \vec{TT^s}, 0.05 \right) \right]}
    \end{aligned}
\end{equation}

\begin{lstlisting}[float, style=mystyle, caption = {Interpolation for signal entrance and exit times.}, label = {lst:tt}]
    signal_in_out_times <- predict(
        # Fit trajectory object
        object = trajectory,
        # Entrance/exit distances for all signals as a numeric vector or dataframe
        new_distances = signal_in_out_distances
    )
\end{lstlisting}

Next, to estimate AoG and SF, we interpolate a sequence of time, distance, and speed points for each trip through each signal at a resolution of 0.1 seconds per point. Once again, this can be accomplished using the \texttt{predict()} method, as shown in Code Block \ref{lst:seq}. We use these point sequences to identify individual stop-and-go cycles, considering any point with a speed below 1.34 meters per second (3 mph) to be stopped then identifying successive sequences of points below this cutoff. With the number of stops made by each trip ahead of each signal, we count the number of trips making at least one stop $N_\text{stopped}^s$ and the number making more than one stop $N_\text{multiple stops}^s$. With a total of $N^s$ trips, the arrival-on-green and portion of split failures can be estimated as follows:

\begin{equation}
    \label{eq:aog}
    \begin{aligned}
        \text{AoG}^s = \frac{N^s - N_{\text{stopped}}^s}{N^s}, \quad \text{SF}^s = \frac{N_{\text{multiple stops}}^s}{N^s}
    \end{aligned}
\end{equation}

\begin{lstlisting}[float, style=mystyle, caption = {Interpolation of sequences of points through each signal.}, label = {lst:seq}]
    trip_interpolated_sequence <- predict(
        # Fit trajectory object
        object = trajectory,
        # Entrance/exit distances for one signal at a time
        distance_lims = signal_in_out_distances,
        # Provide timestep -- the temporal resolution of the sequence for each trip
        timestep = 0.1,
        # Derivative values to return -- 0 for distance, 1 for speed
        deriv = c(0, 1)
    )
\end{lstlisting}

\begin{figure}[ht]
    \centering
    \includegraphics[width=\textwidth, keepaspectratio]{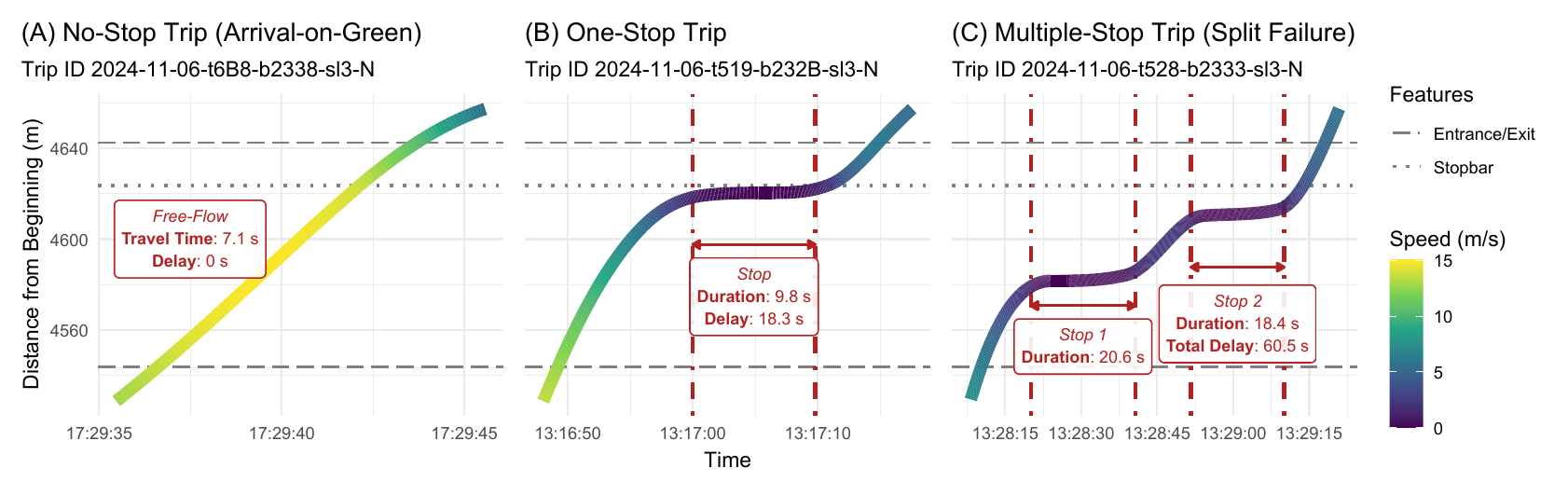}
    \caption{Example IndyGo Red Line 90 NB trajectories at Shelby \& Morris showing (A) a free-flow trip arriving on green, (B) a trip making a single stop, and (C) a split failure making multiple stops.}
    \label{fig:ex_trajs}
\end{figure}

Example trajectories are shown in Figure \ref{fig:ex_trajs}. In (A), the trip likely arrives at Shelby \& Morris on a green and traverses the signal at free-flow. In (B), the trip makes one stop at the signal stopbar; between the deceleration, stop, and acceleration, the total delay is 18.3 seconds. Finally, the trip in (C) makes two distinct stops, indicating a likely split failure, with a total delay of 60.5 seconds. To demonstrate \texttt{transittraj}'s visualization capabilities, we have animated the same three trajectories through space using \texttt{plot\_animated\_map()}, and along a simplified straight-line route using \texttt{plot\_animated\_line()}. These are available as Videos \ref{vid:anim_map} and \ref{vid:anim_line}, respectively, in the Supplementary Materials.

As with the cleaning workflow, we measure the performance of the \texttt{predict()} method by repeating both calls 20 times and reporting the median processing time required by each.

\subsection{Results}

The results of the data cleaning process are shown in Table \ref{tab:cleaning_results}. Through the cleaning process, roughly 23.6\% of observations and 4.8\% of trips were removed; steps 1 (buffer) and 6 (removing insufficient trips) had the largest effects. The exact number of points and trips effected will depend on the user's chosen decisions variables. The sum of the median processing times is 191.4 seconds (3.18 minutes).

\begin{table}[ht]
    \centering
    \caption{AVL Cleaning Results}
    \label{tab:cleaning_results}
    \begin{tabularx}{\textwidth}{W YY Z Y Z}
        \toprule
        Step & Number of Observations & Change in Observations & Number of Trips & Change in Trips & Median Time (s) \\
        \midrule
        Initial &	1,707,095 &	- &	3,256 &	- & - \\
        1 \& 2  &	1,348,250 &	-358,845 (-21.0\%) & 3,254 & -2 (-0.1\%) & 23.6 \\
        3 &	1,345,613 &	-2,637 (-0.2\%) & 3,246 & -8 (-0.2\%)   & 0.9 \\
        4 &	1,344,129 &	-1,484 (-0.1\%) & 3,246 & -0 (-0.0\%)   & 55.6 \\
        5 &	1,335,074 &	-9,055 (-0.7\%) & 3,241 & -5 (-0.2\%)   & 0.4 \\
        6 &	1,304,499 &	-30,575 (-2.3\%)& 3,100 & -141 (-4.4\%) & 1.1 \\
        7 & 1,304,499 &  -0 (-0.0\%)    & 3,100 & -0 (-0.0\%)   & 73.3 \\
        Curve & -     &  -              & 3,100 & -0 (-0.0\%)   & 36.5 \\
        \bottomrule
    \end{tabularx}
\end{table}

\begin{figure}[ht]
    \centering
    \includegraphics[width=\textwidth, keepaspectratio]{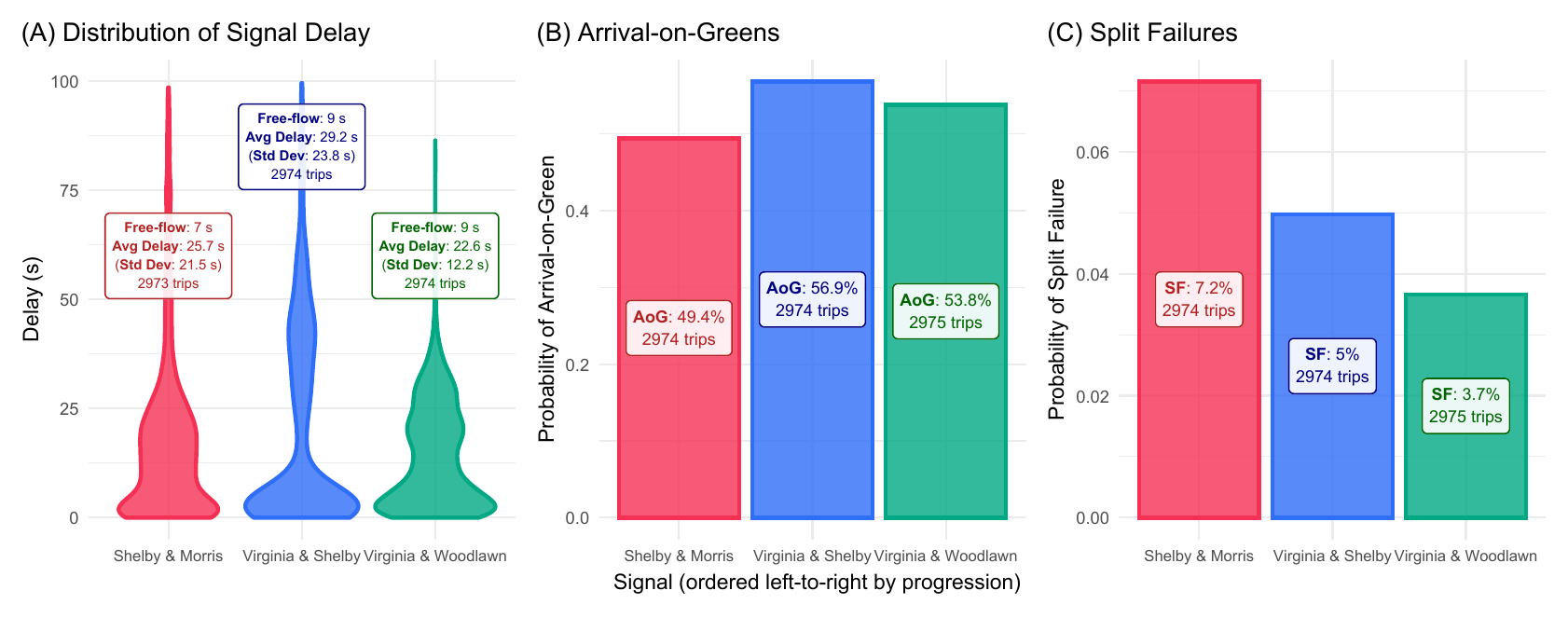}
    \caption{Aggregated IndyGo Red Line 90 NB signal performance metrics through Fountain Square.}
    \label{fig:tspm_res}
\end{figure}

Next, interpolation for delay calculations (Code Block \ref{lst:tt}) took a median of 3.4 seconds for all trips through all three signals, and interpolation for AoG and SF calculations (Code Block \ref{lst:seq}) took a median of 72.4 seconds for all trips through all three signals. Results of these analyses are shown in Figure \ref{fig:tspm_res}. In (A), violin plots show the distribution of delays at each signal, with the largest average delay at Virginia \& Shelby. Despite having the best AoG on the corridor, the violin plot shows that delayed vehicles at Virginia \& Shelby tend to stop for longer than vehicles at the other two signals. Next, (B) shows the AoG of each signal. As expected for a coordinated corridor, the first signal in the progression has a low AoG, followed by higher AoGs at downstream signals. Finally, (C) shows the probability of a transit vehicle experiencing a split failure at each signal. This probability is high at Shelby \& Morris, at 7.2\%, but falls substantially downstream.

As our goal here is simply to demonstrate the derivation of common performance measures using \texttt{transittraj}, we do not explore them further here; we point the reader to \citet{jackson_use_2024} for a more thorough discussion of how signal performance metrics may be used to evaluate TSP systems. This case study does show, however, how the high-resolution AVL data that many agencies already have can be used to easily estimate custom microscopic, between-stop performance indicators; conventionally, signal performance metrics are derived through signal controller data, a difficult source for many transit agencies and researchers to access. Future research may explore comparisons between signal metrics estimated using transit AVL trajectories and those estimated using the conventional data sources. Practitioners seeking to gain a more complete view of progression along their own corridors should consider both directions of travel, time of day differences, and the corridor's coordination plan.

\section{Robustness Check}
\label{sec:robustness}

To quantify the error in reconstructed trajectories across various polling frequencies and interpolation methodologies, we utilize a cross-validation (CV)-style approach. Here, we simulate low-frequency AVL datasets by removing portions of the original AVL stream, then apply the \texttt{transittraj} workflow to fit trajectories to this low-frequency dataset and calculate the trajectories' error relative to the original AVL pings. As these pings are themselves affected by random GPS noise, the errors calculated here are not relative to the ground-truth location of each bus; unfortunately, a ground-truth vehicle location dataset is not available to us. Despite this limitation, we believe that this test can inform analysts on the AVL polling frequencies required to reconstruct high-quality trajectories, and provides them with rough estimates of the magnitude of error to expect. The following sub-sections describe the methods and results of this robustness check. While we do not have permission to share the raw data, we have made available all code used to perform this check through a supplementary GitHub repository (\url{https://github.com/UTEL-UIUC/transittraj_paper-supp}).

\subsection{Methods}

\begin{figure}[ht]
    \centering
    \includegraphics[width=\textwidth, keepaspectratio]{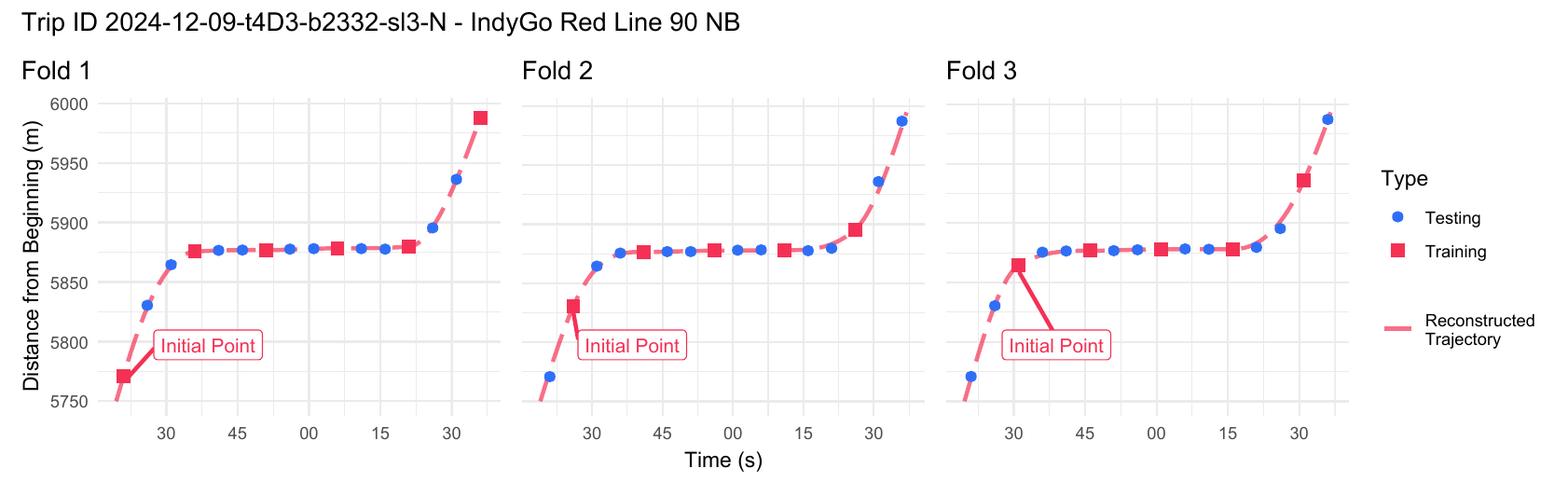}
    \caption{Example CV at 15-second simulated frequency with 5-second raw pings (3 folds). Training points are shown as red squares, with the VCHIP-ME trajectory reconstructed from those training points shown underneath.}
    \label{fig:ex_traj_CV}
\end{figure}

We evaluate the proposed workflow on data with simulated polling frequencies $f_\text{sim}$ ranging from 10 to 60 seconds, in 5-second increments (i.e., $f_\text{sim} \in \left\{ 10, 15, \dots, 55, 60 \right\}$). We begin by selecting all trips on the northbound Red Line with average polling frequencies under 7 seconds (corresponding to 5-second intervals over most of the trip, with random gaps) and remove points which are off-route (step 1), an overlapped sub-trip (step 3), or a trip tail (step 5). In total, 1,162 trips meet these criteria. Next, for each simulated frequency, we select a training dataset by keeping every $f_\text{sim} / 5$ points. All remaining points are used as the testing dataset. This is repeated $f_\text{sim} / 5$ times, shifting the training dataset by one point each time. An example for $f_\text{sim} = 15$ seconds, with $15 / 5 = 3$ folds, is shown in Figure \ref{fig:ex_traj_CV}.

For each CV fold, we apply the same cleaning methodology discussed in Section \ref{sec:case_study_methods}. To understand the relative performance of the different trajectory reconstruction methods, we then apply three separate interpolation methods: first, the recommended VCHIP-ME cubic splines as proposed by \citet{robbennolt_comparative_2026}; second, PCHIP-ME, cubic splines with monotonic enforcement (step 7, Section \ref{sec:mono}) but without observed bus speeds, an approach explored by \citet{huang_reconstructing_2023} and \citet{robbennolt_comparative_2026}; and third, simple linear interpolation without monotonic enforcement (skipping step 7) and without knowledge of observed bus speeds.


Once each trajectory was fit to each training dataset, we calculate two error terms relative to the raw testing dataset: first, the one-dimensional distance along the route between each test point and the fit trajectory; and second, the difference in speed between each test point and the trajectory. Interpolated speeds for VCHIP-ME and PCHIP-ME can be retrieved directly from the interpolating polynomials; for the linear trajectories, we estimate speeds using the slope between each testing point's nearest leading and lagging training points. We then calculate the root mean square error (RMSE) of each of each error type across all folds of each trajectory type and simulated frequency. Finally, to identify spatial patterns in error, we group point errors into 500-meter segments along the route, and calculate the distance and speed RMSEs within each zone.

\subsection{Results}

\begin{figure}[ht]
    \centering
    \includegraphics[width=\textwidth, keepaspectratio]{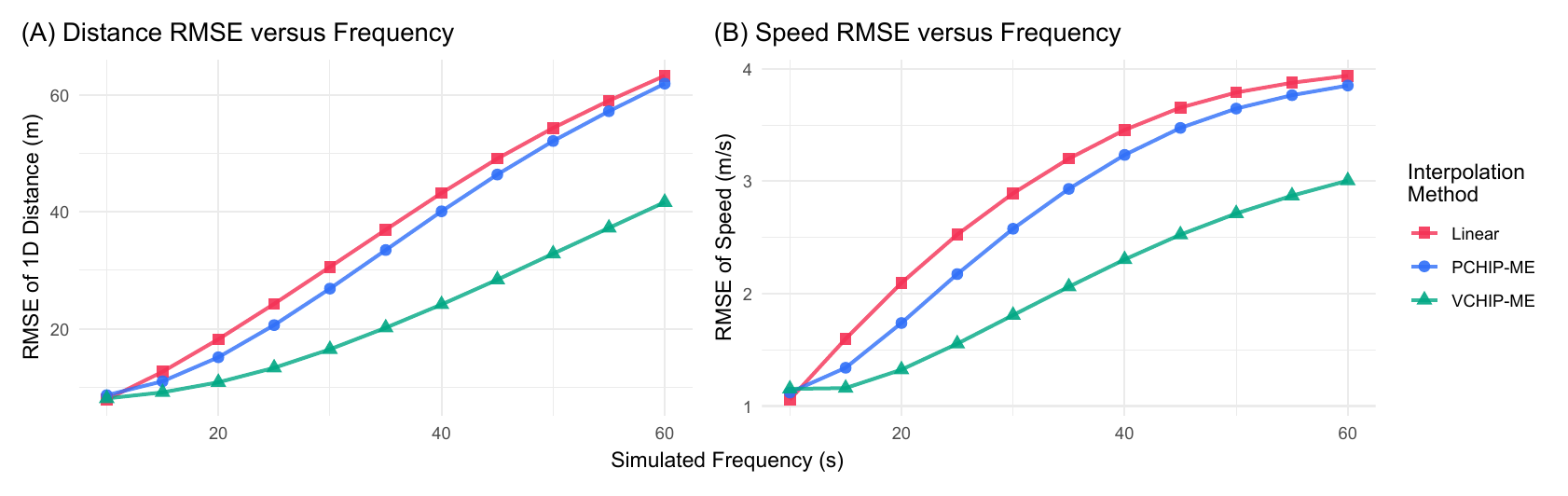}
    \caption{Speed and distance RMSEs versus polling frequency across all trajectory reconstruction types.}
    \label{fig:res_scale}
\end{figure}

Table \ref{tab:CV_results_2} summarizes the route-wide distance and speed RMSEs across simulated polling frequencies $f_\text{sim} \le 30$ seconds and all trajectory reconstruction methods. VCHIP-ME tends to outperform both simple linear interpolation and velocity-ignorant PCHIP-ME reconstruction techniques, corroborating the findings of \citet{robbennolt_comparative_2026}. The exception is $f_{\text{sim}} = 10$ seconds, where the distance and speed RMSE between the trajectories and raw data is slightly lower for linear interpolation than VCHIP-ME. For $f_{\text{sim}} > 10$ seconds, the VCHIP-ME far outperforms the other two methods, and PCHIP-ME tends to perform only slightly better than simple linear interpolation. Figure \ref{fig:res_scale} visualizes how the RMSE estimates vary with polling frequency for each trajectory type. Distance RMSEs appear to increase roughly linearly with polling frequency, with PCHIP-ME and linear trajectories increasing at a faster rate than VCHIP-ME. Speed RMSEs appear concave with respect to frequency, plateauing as frequencies approach 60 seconds. Among VCHIP-ME trajectories, error stays relatively consistent for $f_{\text{sim}} \le 20$ seconds, with the RMSE being no larger than roughly 10.9 meters in distance and 1.3 m/s in speed.


\begin{table}[ht]
    \centering
    \caption{RMSE of Reconstructed Trajectories Relative to Raw 5-second Test Pings}
    \label{tab:CV_results_2}
    \begin{tabularx}{\textwidth}{YYYYY}
        \toprule
{Simulated Frequency (sec)}                    & Number of Trips          & Method   & RMSE of Distance (m) & RMSE of Speed (m/s) \\ \midrule
\multirow[t]{3}{*}{10}                         & \multirow[t]{3}{*}{1162} & Linear   & 7.91                    & 1.06                \\
                                               &                          & PCHIP-ME & 8.66                    & 1.12                \\
                                               &                          & VCHIP-ME & 8.10                    & 1.15                \\
\multirow[t]{3}{*}{15}                         & \multirow[t]{3}{*}{1159} & Linear   & 12.67                   & 1.60                \\
                                               &                          & PCHIP-ME & 11.03                   & 1.34                \\
                                               &                          & VCHIP-ME & 9.14                    & 1.16                \\
\multirow[t]{3}{*}{20}                         & \multirow[t]{3}{*}{1147} & Linear   & 18.21                   & 2.09                \\
                                               &                          & PCHIP-ME & 15.13                   & 1.74                \\
                                               &                          & VCHIP-ME & 10.86                   & 1.32                \\
\multirow[t]{3}{*}{25}                         & \multirow[t]{3}{*}{1141} & Linear   & 24.27                   & 2.53                \\
                                               &                          & PCHIP-ME & 20.63                   & 2.17                \\
                                               &                          & VCHIP-ME & 13.33                   & 1.55                \\
\multirow[t]{3}{*}{30}                         & \multirow[t]{3}{*}{1127} & Linear   & 30.58                   & 2.89                \\
                                               &                          & PCHIP-ME & 26.87                   & 2.58                \\
                                               &                          & VCHIP-ME & 16.50                   & 1.81                \\
\bottomrule
    \end{tabularx}
\end{table}

\begin{figure}[ht]
    \centering
    \includegraphics[width=\textwidth, keepaspectratio]{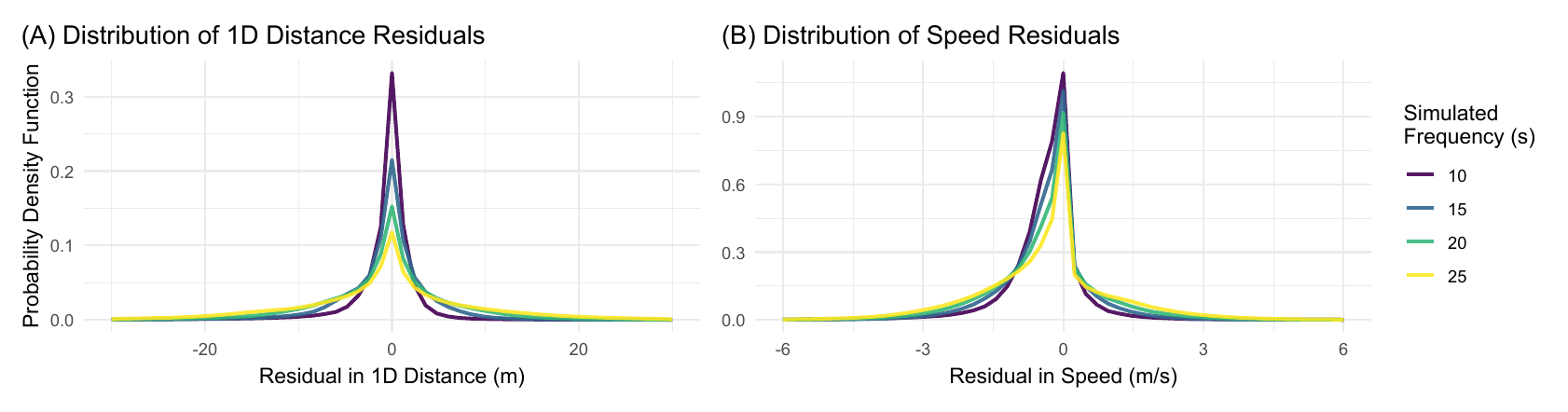}
    \caption{Distribution of residuals in distance and speed estimates for VCHIP-ME trajectories.}
    \label{fig:CV_resid}
\end{figure}

Figure \ref{fig:CV_resid} plots the distribution of the residuals for distances and speeds for VCHIP-ME trajectories at 10 through 25 second frequencies. The distribution of distance residuals is symmetric and centered on zero, indicating that the trajectories may be an unbiased estimate of the projected GPS pings, which themselves are unbiased estimates of the true distance traveled \citep{punzo_assessment_2011}. The speed residuals appear to have a slight negative bias, though the variance is tight.

\begin{figure}[p]
    \centering
    \includegraphics[width=\textwidth, height=\textheight, keepaspectratio]{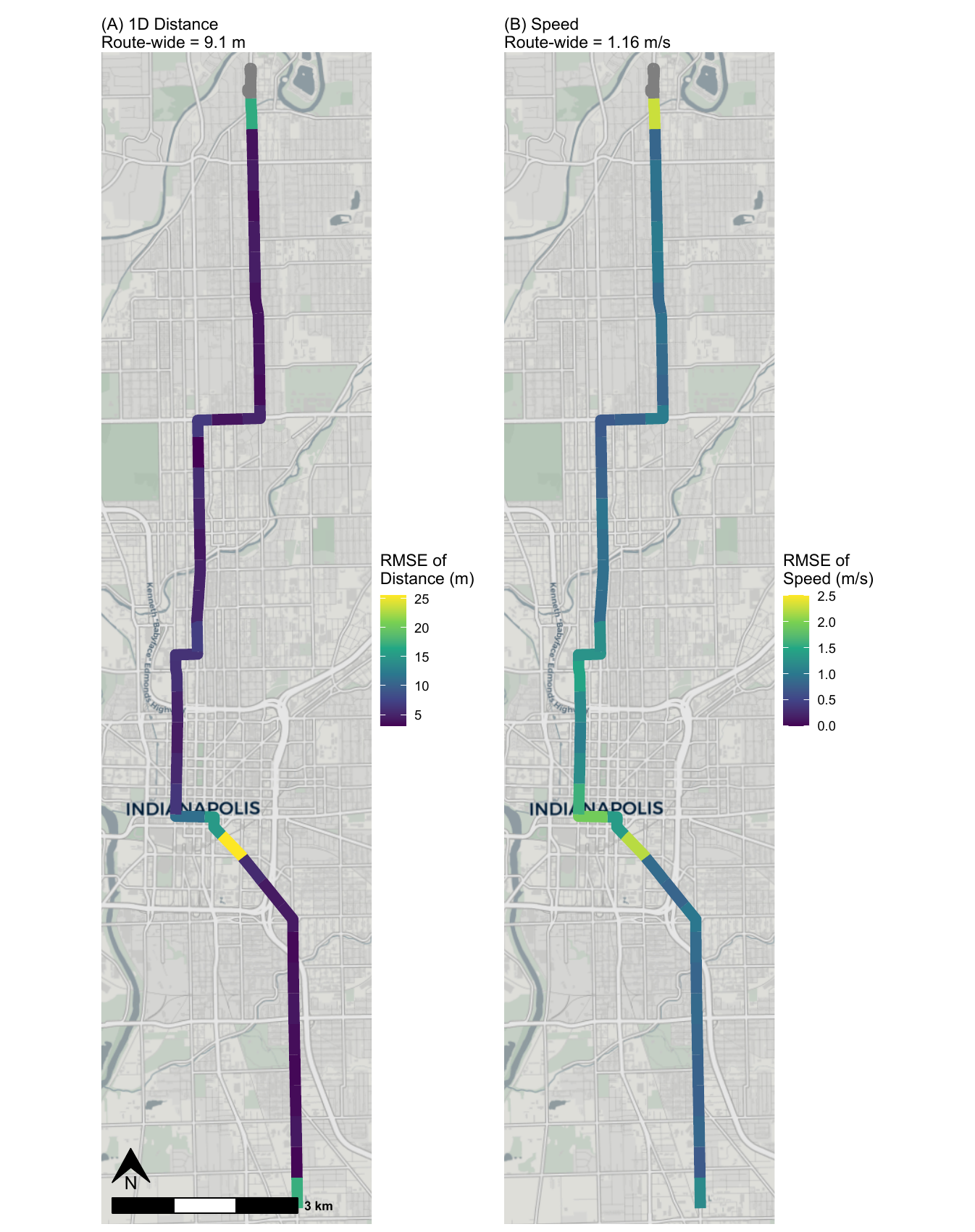}
    \caption{RMSE of 15-second reconstructed VCHIP-ME trajectories across 500-meter segments along the IndyGo Red Line 90 NB.}
    \label{fig:map_CV}
\end{figure}

Finally, Figure \ref{fig:map_CV} shows the spatially-pooled VCHIP-ME RMSEs for 15-second simulated frequencies and indicates that the spatial distribution of these errors is heterogenous. The reconstructed trajectories tend to deviate most from the raw data through downtown Indianapolis, where urban canyon effects may be stronger, and at terminals, where off-route operations may create unexpected shapes once projected onto the route. At $f_\text{sim} = 15$ seconds, the segment with the largest error is just south of downtown, with a distance RMSE of 25.5 meters. Here, the roadway passes underneath a rail viaduct and parking garage, likely causing larger errors in the GPS readings. Most segments have less error, with 25 of 39 (64\%) having a distance RMSE below 5 meters, and 27 (69\%) having a speed RMSE below 1 m/s.

In summary, the proposed \texttt{transittraj} workflow appears to provide stable and low-error trajectories when AVL polling frequencies are 20 seconds or better. Beyond 30 second frequencies, the quality reconstructed distances and speeds falls drastically. In line with previous research, we find that VCHIP-ME trajectories substantially outperform both simple linear interpolation and speed-ignorant PCHIP. While \texttt{transittraj} supports all three methods, we recommend practitioners and researchers use VCHIP-ME whenever possible. The workflow performs best when vehicles are cruising on their mainline through low- to medium-density development; unpredictable terminal operations and dense urban canyons can cause larger errors in reconstructed distances and speeds. It is worth noting, once again, that the residuals and RMSEs estimated here are relative to the 5-second GPS pings; future research should explore techniques for validating trajectories against ground-truth position data.

\section{Conclusion}
\label{sec:conclusion}

In this paper, we propose a methodology for cleaning noise and blunders in transit AVL data, and provide an open-source tool, \texttt{transittraj}, to implement these methods and reconstruct continuous, monotonic, differentiable, and invertible vehicle trajectories. We demonstrate the package on a large, real-world dataset from IndyGo with over 1.7 million points and 3,000 trips. \texttt{transittraj} is efficient, cleaning the data and reconstructing trajectories in just over 3 minutes on a standard consumer laptop. Once trajectories have been reconstructed, custom microscopic performance metrics can easily and quickly be formulated using a single function call; we demonstrate this by estimating various traffic signal performance indicators, measures which are normally estimated using difficult-to-access signal controller logs. At AVL polling frequencies of 15 seconds or better, the reconstructed VCHIP-ME trajectories have low error relative to the raw data, with unbiased distance estimates and an RMSE under 10 meters. Practitioners using \texttt{transittraj} should expect accurate and precise reconstructions with polling frequencies of 20 seconds or better, with higher error in urban canyons and underpasses.

\texttt{transittraj} has a handful of limitations, opening directions for future work or community contributions. First, the cleaning steps which are recursive (step 7, monotonic enforcement) or require a sliding window (step 4, outlier detection) remain slow, and could likely by sped up substantially if their back-ends were re-written in a lower-level language like \texttt{C++} \citep[Chapter 25]{wickham_advanced_2019}. Additionally, \texttt{transittraj} was designed to work on a single route shape, limiting network-level analyses; future versions of the workflow should explore a generalized approach, possibly through map-matching techniques.

Beyond the proposed package itself, there are a few additional paths forward for the study of transit vehicle trajectories. First, some agencies are gaining access to higher-frequency AVL pings (for example, \citet{coghlan_assigning_2019} use AVL with 1-second polling), where smoothing or state estimation techniques may out-perform interpolation \citep{toledo_estimation_2007}. Future work should evaluate additional reconstructing techniques at these frequencies and compare them to the methods explored here. When doing so, researchers should also explore additional sources of ground-truth trajectory data, especially between scheduled stops. Finally, future work may compare the trajectory-centered approach to signal performance metrics used here to traditional methodologies reliant on signal controller logs \citep{nevers_nchrp_2020,jackson_use_2024}.

\newpage
\section*{Supplementary Material}

While we do not have permission to share the raw data from IndyGo, we have made available all code used to demonstrate, evaluate, and visualize \texttt{transittraj}'s workflow in this paper. This is available at our supplementary GitHub repository: \url{https://github.com/UTEL-UIUC/transittraj_paper-supp}. As an open-source tool, the \texttt{transittraj} source code is also available on the project's primary GitHub repository: \url{https://github.com/UTEL-UIUC/transittraj}.

\begin{video}[ht]
    \centering
    \includegraphics[width=0.18\textwidth, keepaspectratio]{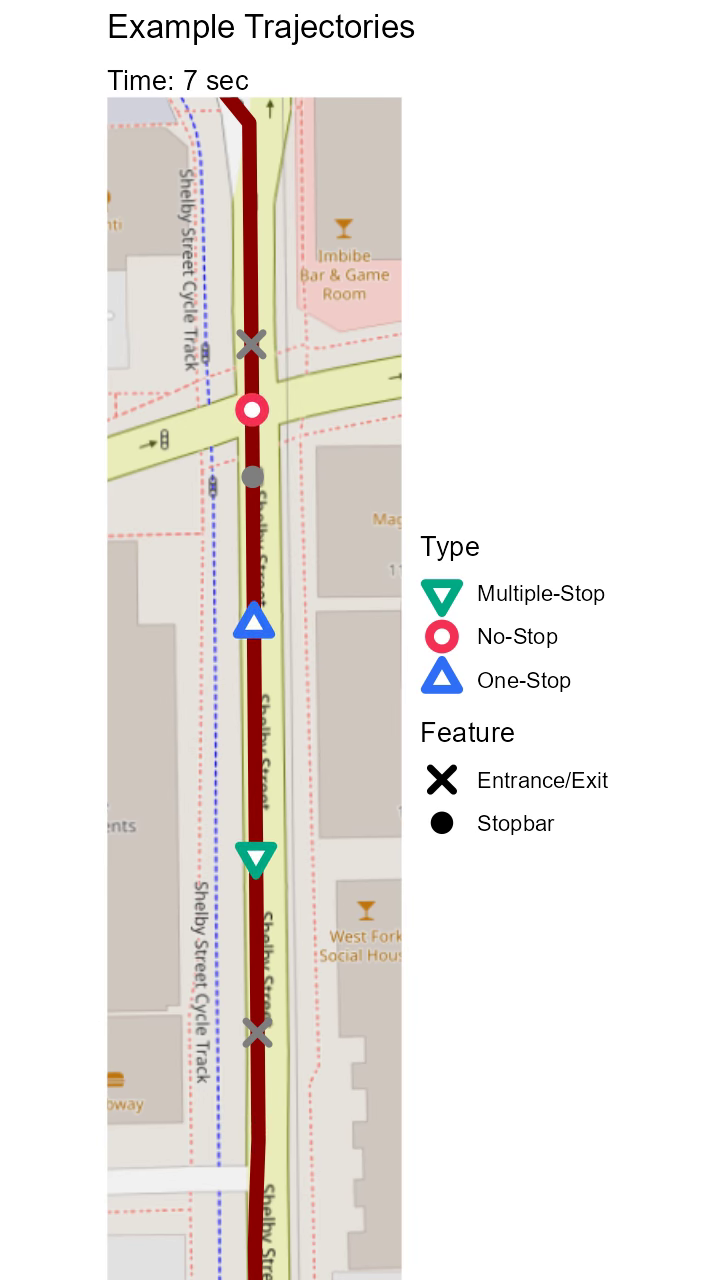}
    \caption{Three vehicles moving through Shelby \& Morris in space with a basemap from OpenStreetMap. The first vehicle does not make any stops, the second makes one stop ahead of the intersection's stopbar, and the third makes two distinct stops ahead the stopbar. These vehicle animations correspond to the trajectories shown in Figure \ref{fig:ex_trajs}. The video is also available online at \url{https://www.youtube.com/watch?v=hCGx4Ki3GRM}.}
    \label{vid:anim_map}
\end{video}

\begin{video}[ht]
    \centering
    \includegraphics[width=0.18\textwidth, keepaspectratio]{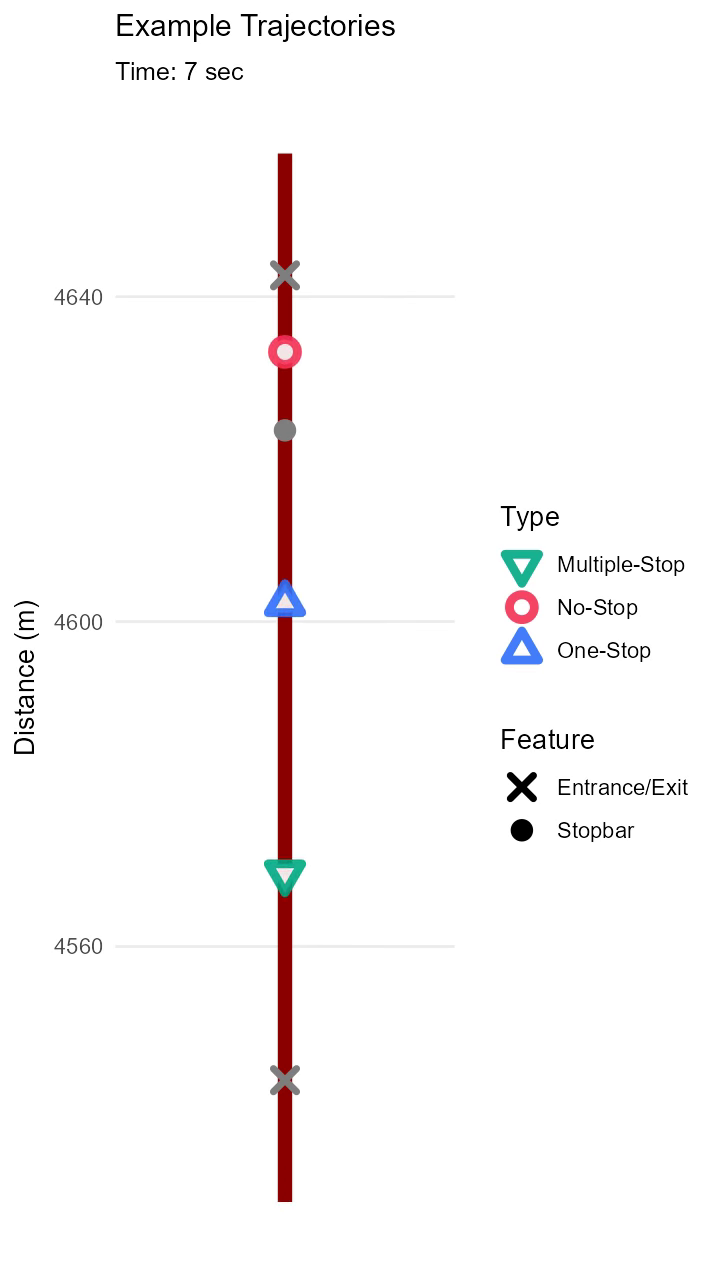}
    \caption{Three vehicles moving through Shelby \& Morris along a simplified straight-line route. The first vehicle does not make any stops, the second makes one stop ahead of the intersection's stopbar, and the third makes two distinct stops ahead the stopbar. These vehicle animations correspond to the trajectories shown in Figure \ref{fig:ex_trajs}. The video is also available online at \url{https://www.youtube.com/watch?v=60AkEtYd9O0}.}
    \label{vid:anim_line}
\end{video}

\section*{Funding}

The authors disclosed no financial support for the research, authorship, and/or publication of this article.

\section*{CRediT authorship contribution statement}

\textbf{Benjamin O'Brien:} Conceptualization, Methodology, Software, Formal analysis, Data Curation, Writing - Original Draft, Visualization. \textbf{Lewis Lehe:} Conceptualization, Writing - Review \& Editing, Supervision, Project administration.

\section*{Declaration of Competing Interests}

The authors declare that they have no known competing financial interests or personal relationships that could have appeared to influence the work reported in this paper.

\section*{Acknowledgements}

We thank Matthew Duffy of IndyGo and Kevin Lee of Illumine Transportation for sharing data and providing guidance in our development and application of these tools. GitHub CoPilot was used to aid in the formatting this manuscript.

\newpage
\bibliographystyle{unsrtnat}
\bibliography{references.bib}

\end{document}